\documentclass[preprint]{aastex} 
 
\usepackage[]{xspace} 
\input{boxedeps}
\input{boxedeps.cfg}
\newcommand{\brg}{Br$\gamma$\xspace} 
\def\h2{H$_2$} 
\newcommand{\tts}{T\,Tau\,S\xspace} 
\newcommand{\ttn}{T\,Tau\,N\xspace} 
\newcommand{\mum}{$\mu$m\xspace} 
\def\arcsec{\hbox{$^{\prime\prime}$}\xspace} 
\def\sun{\hbox{$_\odot$}\xspace}
\def\suneq{_\odot}
\def\degr{\hbox{$^\circ$}\xspace}

\shorttitle{Spatially resolved imaging spectroscopy of T Tauri} 
\shortauthors{Kasper et al.} 
 
\begin{document} 
 
\title{Spatially resolved imaging spectroscopy of T Tauri} 
 
 
\author{M. E. Kasper\altaffilmark{1}, M. Feldt, T. M. Herbst, S. Hippler} 
\affil{Max-Planck-Institut f\"{u}r Astronomie, K\"{o}nigstuhl 17,  
69117 Heidelberg, Germany} 
\email{mkasper@eso.org} 
 
\and 
 
\author{T. Ott, L. E. Tacconi-Garman}  
\affil{Max-Planck-Institut f\"{u}r 
extraterrestrische Physik, Garching, Germany} 
 
 
\altaffiltext{1}{now with European Southern Observatory,  
Karl-Schwarzschild-Str.\ 2, 85748 Garching, Germany} 
 
\begin{abstract} 
  The small separation of the individual stellar components 
  in the T÷Tau system has prevented spectroscopy of the 
    individual components in the near-infrared (NIR) in the past.  We 
  present NIR data (H- and K-band) on this object taken using the ALFA 
  adaptive optics (AO) system in combination with the 3D 
  integral field spectrograph.  Except for the Brackett series seen in 
  emission, the NIR spectra of \tts{} appear featureless, suggesting 
  that warm dust dominates the radiation.  No ro-vibrational lines of 
  molecular hydrogen were found in close vicinity to the stars.  The 
  Brackett line emission from \ttn{} and S could not be resolved with 
  the 3.5-m-telescope, implying that it arises within 6 AU from the 
  stars.  The line ratios between Br$\gamma$ (2.166\,\mum{}) and Br10 
  (1.736\,\mum{}) toward \ttn{} and \tts{} are similar, suggesting similar 
  selective extinctions towards the respective emitting regions.  Our 
  results are consistent with a model that describes \tts{} as a 
  pre-main sequence star surrounded by a small edge-on disk, leaving 
  the polar regions relatively unobscured.  We present numerical 
  simulations which support this model. 
\end{abstract} 
 
 

 
\section{Introduction} 
 
T~Tauri is a complex system of multiple (at least three) 
stellar components, jets, and outflows unlike that seen in any other T 
Tauri star.   
 
The optically visible primary, \ttn{}, is an early K star 
\citep{cohen79}, which has a companion at 0.7\arcsec{} separation to 
the south found by speckle interferometry in the NIR \citep{dyck82}.  
This infrared companion, \tts{}, has not yet been detected at visible 
wavelengths \citep{stapelfeldt98}.  \tts{} exhibits a spectral energy 
distribution (SED) suggestive of a very young embedded source with a 
luminosity of about 11 L\sun{}, making it somewhat more luminous than 
\ttn{} \citep{koresko97}.  It shows an apparent lack of cold dust 
emission, implying a rather small mass for a circumstellar disk 
\citep{hoger97}.  \tts{} underwent a brightness flare of two 
magnitudes at various wavelengths between 1989 and 1991 
\citep{ghez91,kobayashi94}.  After a quiescent low brightness phase in 
the mid 90s \citep{herbst97}, another flare started in 1998 
\citep{roddier00} and lasts until today. 
 
By speckle holography observations from 1998, \citet{koresko00} found
a companion to \tts{} at a separation of only 0.05\arcsec{} and
assigned the name \tts{}b to this component.  Only two years later,
this companion was found at a distance of 0.08\arcsec{} with its
position angle changed by 28\degr{} with respect to \tts{}
\citep{koehler00}.  Orbital motion studies of \ttn{} and S using
speckle interferometry \citep{ghez95} and AO \citep{roddier00} imply a
minimum mass of 3.3 M\sun{} for the whole T~Tauri system.
 
The near environment of T~Tauri is a source of surprisingly strong 
extended ro-vibrational emission of molecular hydrogen 
\citep{beckwith78}.  At higher resolution, the spatial distribution of 
this excited \h2{} shows two separate outflows, perpendicular to each 
other in projection, and rather strong emission in close vicinity to 
both stars \citep{herbst96}.  The excitation appears to be produced by 
shock heating rather than UV fluorescence.  It could either be caused 
by infall of material onto the star, or by outflows intercepting 
ambient matter. The \brg{} emission appeared to be confined to \ttn{}.  
In 1996/1997, one year after Herbst's observations, \tts{} showed weak 
\brg{} emission, while \h2{} was barely detectable \citep{beck00}.  
 
Recent optical spectroscopy \citep{solf99} indicated that the outflows
seen in \h2{} are in fact almost perpendicular with the north-south
component seemingly originating from \tts{} at 79\degr{} inclination
from the line of sight, and the east-west component originating from
\ttn{}.  The latter coincides with the HH 155 jet, which is seen at an
inclination angle of 23\degr{} \citep{eisloeffel98}.
 
\section{Observations} 
 
The observations described here were taken with the ALFA adaptive 
optics (AO) system \citep{kasper00} on the 3.5-m telescope at the 
German-Spanish Astronomical Center on Calar Alto.  Using \ttn{} 
($m_{V}=9.6$) as a natural guide star, ALFA delivered angular 
resolutions (FWHM) of 0.14\arcsec{} in H- and 0.16\arcsec{} in K-band 
throughout the observations.  The integral field spectrometer 3D 
yields spectra of an entire $16 \times 16$ pixel field at resolutions 
of $R \approx 1000$ to $R \approx 2100$ ($\lambda / \Delta\lambda$) 
\citep{weitzel96, anders98}.  The image scale of 0.07\arcsec{} per 
pixel results in a field of view (FOV) of  one arcsecond.  Due to the 
small FOV, sky frames have to be observed in addition, effectively 
doubling the observing time. 
 
T~Tauri was observed with ALFA and 3D on September 28, 1999 in K-band
and on October 1, 1999 in H-band with spectral resolutions of $R
\approx 1100$ and $R \approx 2100$ respectively.  The total
integration time on source was 18 and 14 minutes, respectively. 
Standard techniques of sky subtraction and flat fielding removed
background emission, dark current, and pixel-to-pixel gain variations. 
The G2V star HD\,26710 was observed at a similar airmass to T~Tauri
and served as a photometric, spectroscopic, and PSF reference. 
Division by the spectroscopic standard and subsequent multiplication
by the sun's spectrum at the same spectral resolution eliminated
telluric and photospheric features of the G2V star and produced the
correct spectral slope across the band.  The final 3D data structure
consists of an image cube with 600 separate wavelength slices, each
containing spatial information over the FOV. This data cube allows one
either to create line maps by subtracting the continuum next to a line
from the line flux itself, or to extract a spectrum anywhere in the
field.
 
\section{Results} 
 
\subsection{Spectra} 
\begin{figure}[htb]
\HideDisplacementBoxes
\centerline{\BoxedEPSF{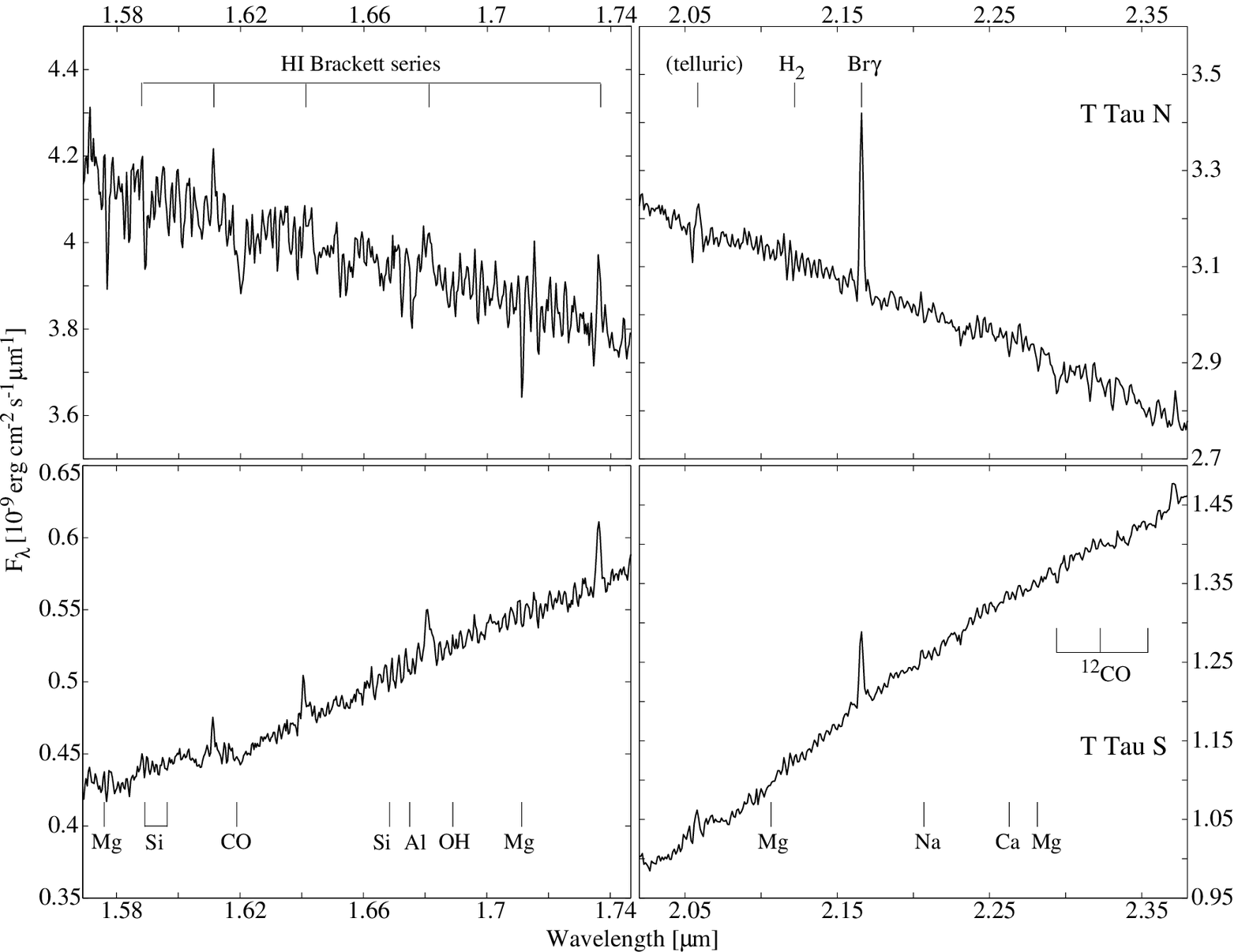}
           }\ForceOff
	   \caption{H- and K-band spectra of \ttn{} and S  
obtained with ALFA and 3D. See text for details of the synthetic 
apertures and photometric calibration. \label{fig-ttspec}} 
\end{figure}

Viewing the 3D data cube signal in a direction perpendicular to the
image planes produces conventional spectra.  The traces in
Figure\,\ref{fig-ttspec} represent the total flux in a 0.25\arcsec{}
square box centered on \ttn{} and \tts{}, respectively.  Results for
\tts{} include also its companion because it was not resolved by our
observations.  The vertical bars and labels denote diagnostic lines in
the actual photometric band.  The detection limit for the lines can be
estimated by measuring equivalent widths (EWs) on the noisy continuum. 
In the case of our spectra, such EWs are well below 0.15\,\AA{},
suggesting that line features with EWs greater than about 0.3\,\AA{}
are detectable at a 2\,$\sigma$ level.
 
Table\,\ref{tab-lineflux} lists the line fluxes and uncertainties 
based on least-squares fits of Gaussian profiles to the lines drawn 
from the spectra. The model includes five parameters: the central 
wavelength $\lambda_{0}$, the amplitude $A$, the standard deviation of 
the Gaussian $\sigma$, and two parameters $m$ and $b$ giving the 
slope and intercept of the linear baseline, respectively.  With the 
model 
\begin{equation} 
        F_{\lambda} = b + m\lambda + A  
        \exp{-\frac{(\lambda-\lambda_{0})^{2}}{2\sigma^{2}}}, 
\end{equation} 
the flux is given by $\sqrt{2\pi} A \sigma$, and the formal 
uncertainty is $\Delta F_{\lambda} = [2 \pi(\sigma^{2} \Delta A^{2} + 
A^{2} \Delta \sigma^{2})]^{1/2}$, where $\Delta A$ and $\Delta \sigma$ 
are those changes in the $A$ and $\sigma$ parameters which increase 
the $\chi^{2}$ per degree of freedom by one \citep{bevington92}.  It 
should be noted that a number of additional effects which 
are not included in the uncertainties have an impact on the line 
measurements.  These effects include, for example, imperfect telluric 
calibration or uncertainty in the brightness of the photometric 
standard. 
 
No significant \h2{} emission could be found within one arcsecond of
the stars.  Brackett recombination lines of atomic hydrogen dominate
the H- and K-band spectra of \ttn{} and \tts{} with comparable EWs for
both stars.  Flux ratios between the \brg{} and Br10 line are about
3.25 for both \ttn{} and \tts{}, consistent with the value expected
for optically thin emission from a gas in local thermal equilibrium
and obeying Menzel Case B recombination \citep[see][p.\
84]{osterbrock89}.
 
Photospheric absorption features as given in Table\,\ref{tab-ttnew} 
are only found for \ttn{}. None are detected in the spectrum of 
\tts{}. The relative strengths of the EWs are consistent with  
those seen in early K spectral standard star spectra  
\citep[e.g.][]{meyer98,wallace97}, but the absolute values are much  
more shallow. This requires a large amount of continuum veiling such  
as the contribution of the hot inner part of a circumstellar disk. In  
the case of \ttn{}, the K-band excess must be of the order 80\,\% of  
the total flux. There is some hint in both stars of the presence of 
the $v = 2-0$ vibrational bandhead of CO near 2.3\,\mum{}. This 
wavelength region suffers from telluric absorption features, however, 
and is difficult to correct completely. Also, there is no evidence of 
additional $\Delta v = 2$ bandheads ($3-1$, $4-2$, etc.) at longer 
wavelengths, which accompany the $2-0$ bandhead in spectra of 
late type stars \citep{kleinmann86}.
 
\subsection{Linemaps}  
\begin{figure}[htb]
\HideDisplacementBoxes
\centerline{\BoxedEPSF{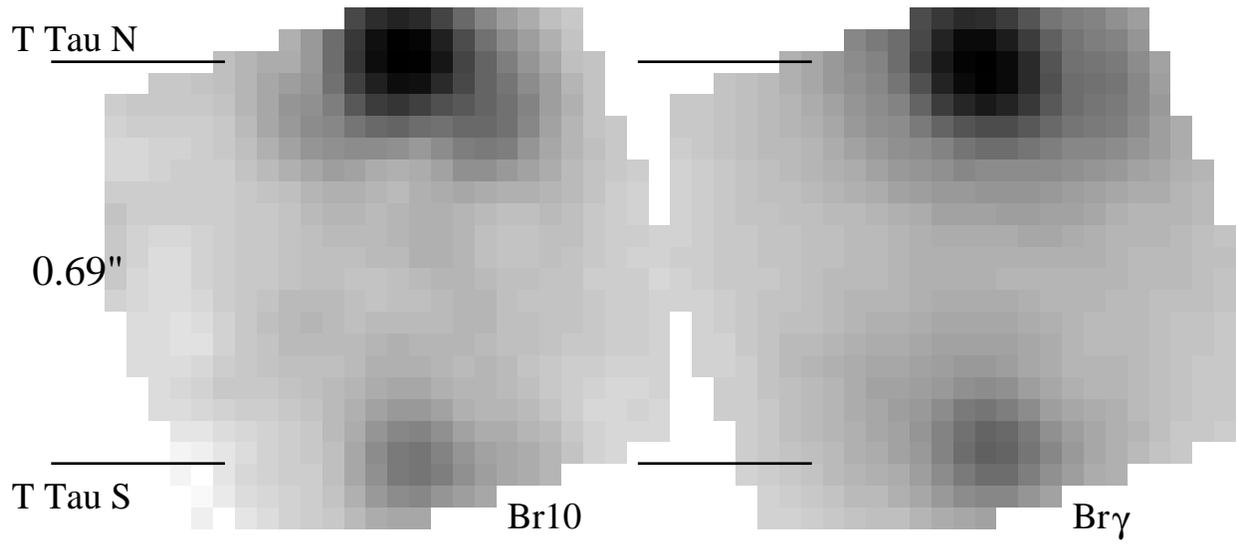}
           }\ForceOff
	   \caption{Continuum subtracted line maps in
Br10 and \brg{}.  The images display the whole FOV of 
3D. A squareroot scaling was chosen to enhance the fainter \tts{}.  
\label{fig-ttbrackettlm}} 
\end{figure}

The continuum subtracted line maps displayed in 
Figure\,\ref{fig-ttbrackettlm} allow us to search for extended Brackett 
emission when compared to images in the adjacent continuum.  The 
spatial resolution of the H-band line maps from October 1999 is about 
0.14\arcsec{} FWHM. When compared to the presumably point-like 
continuum PSF, a potential source extension in a Brackett line image 
would add to this value.  We performed empirical tests 
using Gaussian fits to the PSF cores in the continuum linemaps next to 
the Br10 line.  We found that, at the distance of T Tau ($\approx$ 
140\,pc), a source extension of more than 6\,AU convolved with the 
mean PSF is required to increase the FWHM significantly ($3 \sigma$) 
compared to the scattering in the continuum PSF FWHMs. The continuum 
subtracted Br10 PSF does not show a significantly larger FWHM than the 
continuum PSFs, meaning that the radiation arises within 6\,AU from 
the stars. Using a similar test on the position showed that the 
Brackett emitting region is coincident with the star at a level of 
0.8\,AU.
 
\section{Discussion \label{sec:discuss}} 
Apart from the Brackett emission lines, the H- and K-band spectra of
\tts{} are featureless.  In comparison to our detection limit of about
0.3\,\AA{}, typical EWs of atomic features for late type standard
stars are of the order of a few Angstroms
\citep{meyer98,wallace97}.  This implies that the stellar photosphere
of \tts{} is heavily veiled, if visible at all, meaning that we
basically observe hot dust.  A scenario, where \tts{} is a star
reddened by large amounts of extinction from cold dust, can thus be
ruled out.
 
Our data do not show any \h2{} emission within one arcsecond of \ttn{}
and S. This is a somewhat surprising result, because \citet{herbst96}
found strong \h2{} emission centered at the positions of \ttn{} and
\tts{}, and \citet{beck00} detected weak \h2{} emission from \tts{}
only.  Brackett recombination lines instead are prominent in the
spectra of both stars, with the contribution of \tts{} being stronger
than observed by \citet{herbst96} who did not detect \brg{} emission
and \citet{beck00} who found a lineflux of $1.6 \pm 0.5 \times
10^{-13}$ erg cm$^{-2}$ s$^{-1}$.  Our observations date from a flare
period, while the two others were carried out during a low brightness
phase of \tts{}.  Hence, the variable emission line luminosity of
\tts{} suggests a common cause with its variable brightness.
 
In an attempt to estimate the excess luminosity produced by an 
accretion event from the \brg{} luminosity, we used the following empirical 
formula given by \citet{muzerolle98}
\begin{equation}
	\log(L_{acc}/L\suneq) = (1.26 \pm 0.19) \log(L_{\rm Br\gamma}/L\suneq) 
	+ (4.43 \pm 0.79),
	\label{eq-muzerolle}
\end{equation}
to derive a value of $L_{acc} = 0.44 \pm 0.09 L\suneq$ for \tts{}. 
The uncertainty includes our measurement error as well as those
related to Equation\,\ref{eq-muzerolle}.  The derived value is
certainly not sufficient to explain brightness flares of \tts{} of
more than one magnitude when taking into account its total luminosity
of more than 10\,L\sun{} \citep{koresko97}.  The larger \brg{} flux
during the brightness flare is however consistent with a foreground
extinction that has decreased from $A_{V} \approx 9$ to $A_{V} \approx
6$ in the transition between quiescent and elevated flux state
\citep{beck00}.  These extinction values were derived from observed
water ice absorption and therefore represent foreground extinction
produced by relatively cold material.  They do not represent the
effects of warm material close to the star.

The extinction towards a Brackett emitting region can be determined by
the observed \brg{} to Br10 flux ratio $x$, if the intrinsic
recombination line flux ratio $x_{0}$ is known.  Although our observed
flux ratio of $x = 3.23$ is consistent with Menzel Case B, this could
be a coincidence since \citet{evans87} did not find the corresponding
hydrogen recombination line ratios in his sample of pre-main-sequence
objects.  Moreover, \citet{natta88} showed that relative intensities
of hydrogen recombination lines created in a mostly neutral stellar
wind, instead of a fully ionized region, are a complex function of
wind parameters and stellar radiation field.  Let us, however, suppose
that the intrinsic recombination line ratios of \ttn{} and \tts{} are
similar, neglecting the actual generation process.  Since the
extinction towards \ttn{} is negligible in the NIR (\citet{cohen79},
$A_{V} = 1.39$), an intrinsic $x_{0} = 3.23$ is adopted.  With the NIR
extinction law from \citet{mathis90}, $A(\lambda) \propto
\lambda^{-1.7}$, it follows that $A_{\rm Br10} / A_{\rm Br\gamma} =
1.46$ or $A_{\rm Br10} - A_{\rm Br\gamma} = 0.46 A_{\rm Br\gamma}$. 
Hence, the observed line ratio depends on extinction as
\begin{equation}
	x = x_{0} \cdot 10^{-0.4(0.46 A_{\rm Br\gamma})}.
	\label{eq-extinction}
\end{equation}
The maximum flux ratio for \tts consistent with the accuracy of our
observations (see Table\,\ref{tab-lineflux}) is $x = 4.1$, yielding an
upper limit on the extinction towards the Brackett emitting region of
$A_{V} \approx 6$.  The star instead must be must be obscured by at least
$A_{V} = 7$ \citep{stapelfeldt98}. The conclusion is that in
contrast to the star, the \brg{} emitting region around \tts{} does
not seem to be heavily obscured.
 
Although, we observed at the diffraction limit of the 3.5-m telescope,
the Brackett emitting region around \ttn{} and S could not be
resolved.  This is consistent with the neutral wind model of
\citet{natta88} that predicts most of the hydrogen recombination line
flux to be produced within a few tenths of an AU around the central
star, a scale not resolved by our data.  This result does not
contradict radio observations showing \tts{} as an extended source of
two circularly polarized lobes approximately 0.15 arcseconds apart
\citep{ray97}, because it is unlikely that this radio emission is
produced by free-free transitions.  Almost certainly, its origin is
gyrosynchrotron radiation which would not be detectable in our
Brackett linemaps.
 
\section{Model calculations} 
A proto-stellar or Class\,I model with an infalling envelope seems 
unlikely for \tts{}, mostly because of a lack of cold dust in its 
immediate environment and the presence of its close companion.  Based 
on the orientation of \tts{}'s jet almost perpendicular to the line of 
sight and the proximity to \tts{}b, \citet{koresko00} proposed a 
scenario in which both stars are surrounded by small disks seen nearly 
edge-on.  This could easily provide the required local extinctions 
without large amounts of dust and could further explain the brightness 
variations by changes in a disk which processes stellar photons via a 
combination of extinction and scattering. 
 
What can our data add to this picture?  The lack of photospheric
absorption features deduced from the H- and K-band spectra rules out a
pure extinction scenario.  The relatively low extinction towards the
Brackett-series emitting region indicated by our data further suggests
a geometry, where the disk is observed at an inclination such that it
obscures the star while leaving the line of sight towards the polar
regions comparatively unobscured.  NIR flux could be contributed by
the hot inner parts of the disk itself, if the optical depths in the
disk are not too large.
 
\subsection{The model}
To test the validity of these assumptions, we performed model
calculations using the radiative transfer (RT) code of
\citet{manske98}.  This code is based on a method described in detail
by \citet{menshchikov97}.  The model consists of a small disk with an
opening angle of 30$\degr$ inclined by 14$\degr$, i.e. almost edge-on. 
Hence, the line of sight towards the central star intersects the disk
just beneath its surface.  The model parameters are given in table
\ref{tab:ModParams}.  The code was used in its 2D mode, calculating at
first the temperatures and mean intensities in the disk's midplane and
at its conical surface.  Intermediate temperatures and mean
intensities are computed by interpolation.  Finally, specific
intensities are computed using a set of rays intersecting the disk. 
Based on these specific intensities, the SED is computed.

\subsection{Result}
\begin{figure}[htb]
\HideDisplacementBoxes
\centerline{\ForceWidth{140mm}
            \BoxedEPSF{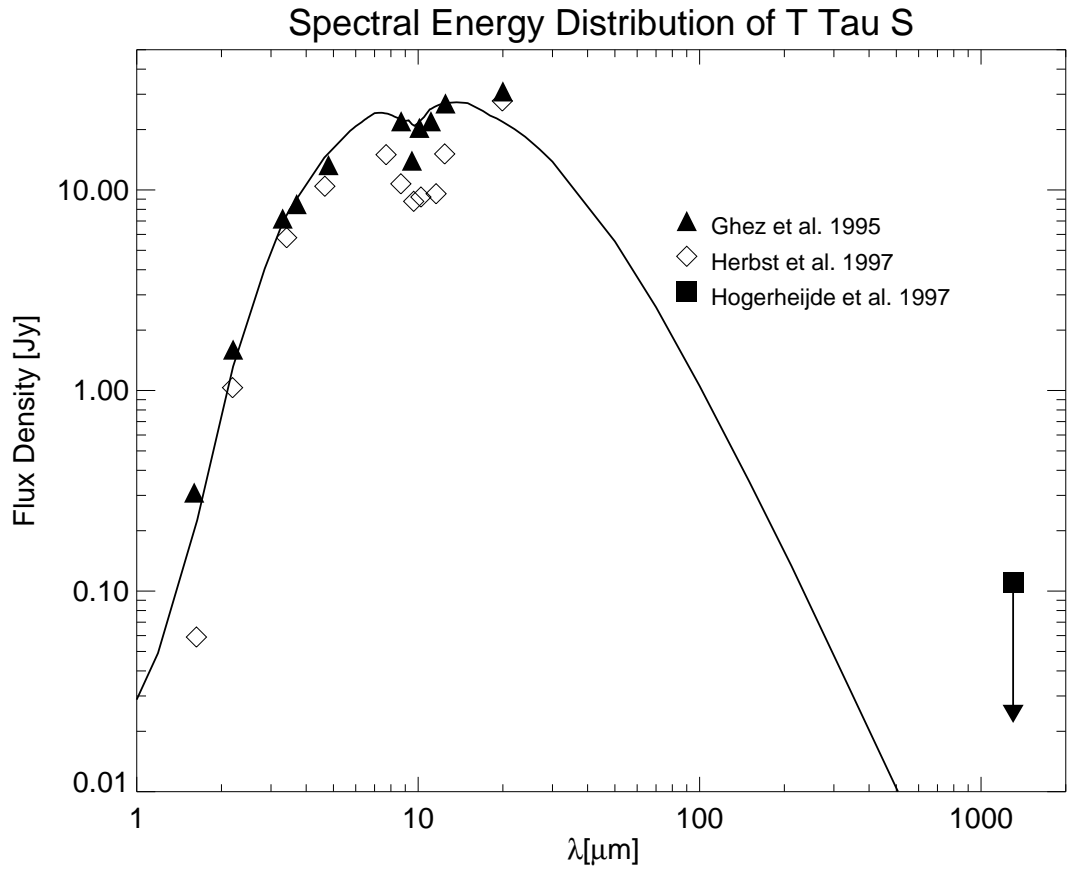}
           }\ForceOff
	   \caption{Near- and mid-infrared photometry of \tts{}. The solid 
line traces the SED of the small edge-on disk model.\label{fig-ttssed}} 
\end{figure}

The SED of \tts as obtained from the model calculations is given in
Figure\,\ref{fig-ttssed}.  To account for the observed foreground
extinction of about $A_V = 6$\,mag during an elevated flux state (see
Sec.\ \ref{sec:discuss}), the result of the RT computation is shown
reddened by this amount using the reddening law from \citet{mathis90}.

The modeled SED well reproduces the flux densities measured by
\citet{ghez95} during the 1990 flare.  Only the data point at
20\,\mum could not be fitted by the small disk, because of a lack of
dust at larger distances to the star and therefore lower temperatures. 
However, the missing 20\,\mum-flux could be produced by the same dust
that provides the foreground extinction.  While little is known about
geometry and temperature distribution of a circumbinary dust torus
around \tts, a rough estimate shows that such a torus located between
25\, and 50\,AU from the stars with a dust density of about
$10^{-14}$\,Kg\,m$^{-3}$ and a temperature of 75\,K can produce the
required foreground extinction and excess flux density of about 10\,Jy
at 20\,\mum.

The very small disk radius of 3\,AU is on the one hand an input
requirement due to the close companion \tts{}b at a projected distance
of about 11\,AU. According to \citet{artymowicz94}, a circumstellar
disk in a binary system has a radius smaller than about 0.45 times the
binary separation for a circular binary orbit.  Larger eccentricities
can reduce this value to 0.2 times the semimajor axis of the binary. 
On the other hand, the small disk radius is required by the RT
calculations, because larger disks tend to produce more MIR and less
NIR flux when observed edge-on and therefore could not fit the data.
  
In conclusion, the concept of a small truncated disk surrounding \tts
with dimensions that fit the close binary system as a circumprimary
disk and produce a high extinction along the line of sight, while
leaving the \brg emitting region (at least the northern half of it)
relatively unobscured, are not contradicted by the RT computations. 
Whether the details of the model are significant, and the above
implications regarding the circumsystem dust torus are correct, could
be subject to further research.
 
\section{Summary} 
 
Using AO compensated imaging spectroscopy, we were able to spatially 
resolve \ttn{} and \tts{} and to obtain individual H- and K-band 
spectra of the two components for the first time.  The observations 
were carried out during an elevated flux state of the variable \tts{}.  
It was not possible to disentangle the contributions of \tts{} and 
\tts{}b. 
 
\h2{} emission was not detected in the immediate vicinity of \ttn{} 
and \tts{}, whereas both exhibit the Brackett series in emission with 
comparable equivalent widths.  Compared to previous observations of 
\tts{} during a quiescent phase, the \brg{} lineflux has increased 
significantly while \h2{} emission could no longer be found.  This is 
evidence for a common cause of variable emission line spectrum and 
luminosity. The accretion luminosities deduced from 
the \brg{} fluxes are typical for pre-main sequence stars and do not 
suggest exceptionally high accretion rates. 
 
The ratio between \brg{} and Br10 luminosity is similar for \ttn{} and 
\tts{}.  Considering the low extinction for \ttn{} and assuming a 
similar intrinsic ratio for both stars, the Brackett series emitting 
region around \tts{} does not appear as heavily obscured as the star.   
 
The Brackett emission lines of \ttn{} and \tts{} arise within a few AU
of the stars.  They were not resolved by our observations.  The
apparent mismatch between Brackett- and radio linemaps are further
evidence against a fully ionized regions as a source of the radio
emission.
 
Photospheric absorption features were only detected for \ttn{} which 
appears heavily veiled.  The absence of any photospheric features in 
the case of \tts{} rules out a pure extinction scenario as an 
explanation for its unusual SED. 
 
Given that we do not see the stellar photosphere of \tts, but do see the 
Brackett series emitting region seemingly unobscured, a small edge-on 
disk may be consistent with the observed SED when viewed at 
an angle such that the star is just obscured.  We carried out 
radiative transfer calculations using a very simple model and found 
that a small edge-on disk can indeed reproduce the SED of 
\tts{}. 
 
\acknowledgments 
 
We are grateful to both the ALFA and 3D teams for their invaluable  
help, and the Calar Alto Observatory for their hospitality. We extend  
our thanks to Tracy Beck for providing a preprint of her paper, and  
to Volker Manske and Olaf Kessel for their support with the radiative  
transfer calculations. We also thank Andy Nelson for discussions on
the mechanics of circumstellar disks in binary systems as well as 
editor and referee for their valuable comments to the manuscript.
 
 
 

 
\clearpage 
 
 
 
\clearpage

\clearpage

\clearpage
 
 
 
\clearpage 
 
\begin{table} 
\begin{center} 
\caption{Emission line fluxes (in $10^{-13}$ erg cm$^{-2}$ s$^{-1}$) 
and EWs (in \AA{}) of \ttn{} and \tts{}.\label{tab-lineflux}} 
\begin{tabular}{llllll}  
        \tableline\tableline 
        Line & Wavelength & \multicolumn{2}{c}{\ttn{}}  
        & \multicolumn{2}{c}{\tts{}}\\ 
        \hspace{1pt} & (\mum{}) & Flux & EW & Flux & EW \\ 
        \tableline 
        H I    & \hspace{1pt} & \hspace{1pt} & \hspace{1pt} & \hspace{1pt} \\ 
        Br13       & $1.611$ & $2.2\pm1.0$ & $0.54\pm0.25$ & $0.2\pm0.05$  
        & $0.45\pm0.1$\\ 
        Br12       & $1.641$ & $0.6\pm0.2$ & $0.16\pm0.03$ & $0.2\pm0.05$  
        & $0.45\pm0.1$\\ 
        Br11       & $1.681$ & $0.9\pm0.6$ & $0.22\pm0.16$ & $0.4\pm0.08$  
        & $0.79\pm0.16$\\ 
        Br10       & $1.736$ & $3.0\pm0.6$ & $0.79\pm0.15$ & $0.8\pm0.09$  
        & $1.42\pm0.16$\\ 
        Br$\gamma$ & $2.166$ & $9.7\pm1.1$ & $3.2\pm0.36$ & $2.6\pm0.29$  
        &$2.16\pm0.25$\\ 
        \h2{}   & \hspace{1pt} & \hspace{1pt} & \hspace{1pt} & \hspace{1pt} \\ 
        1-0 S(1)   & $2.121$ & \ldots\tablenotemark{a}  & \ldots & \ldots & \ldots\\ 
        \tableline\tableline 
\end{tabular} 
\tablenotetext{a}{not detected} 
\end{center} 
\end{table} 
 
\clearpage 
 
\begin{table} 
\begin{center} 
\caption{Equivalent widths of atmospheric features of  
\ttn{}. \label{tab-ttnew}} 
\begin{tabular}{lllllll} 
        \tableline\tableline 
        Species & \multicolumn{3}{l} {Mg} & Al & OH & Ca \\  
         \tableline 
        $\lambda_{0}$\,(\mum{}) & 1.575 & 1.711 & 2.282 & 1.674 & 1.688 & 2.263 \\ 
        EW\,(\AA) & $0.84 \pm 0.45$ & $0.55 \pm 0.21$ & $0.33 \pm 0.12$ &  
        $0.75 \pm 0.44$ & $0.51 \pm 0.33$ & $0.27 \pm 0.18$ \\ 
\end{tabular} 
\end{center} 
\end{table}

\clearpage 
 
\begin{table}
\begin{center}
\caption{Parameters used in the radiative transfer calculation to 
model the SED of \tts{}. 
\label{tab:ModParams}}
\begin{tabular}{rl}
        \tableline\tableline
        \multicolumn{2}{l} {Central object} \\
        \tableline
        Luminosity: & 16 L\sun \\
        Temperature: & 5000\,Kelvin \\
        Distance: & 140\,pc \\
        \tableline
        \multicolumn{2}{l} {Dust Distributions} \\
        \tableline
        Density distr.: & $\sim r^{-2}$ \\
        Outer radius: & 3\,AU \\
        Inner radius: & 0.9\,AU \\
        \tableline
        \multicolumn{2}{l} {Dust Properties} \\
        \tableline
        Size distr.: & $a=0.005-0.25$\,\mum \\
        & $N(a) \sim a^{-3.5}$ \\
        Silicate/Carbon ratio: & Si:C 3:2 \\
        Optical data : & \citet{preibisch93}\\
        & \citet{draine84} \\
        Optical depth of disk: & $\sim$700 at 0.5\,\mum \\
        Total dust mass: & $2.6 \times 10^{-5}$\,M\sun \\
        Foreground extinction: & $A_V=6$\,mag \\
        \tableline
        \multicolumn{2}{l} {Disk geometry}\\
        \tableline
        Opening angle: & $30\deg$ \\
        Inclination: & $14\deg$ \\
\end{tabular}
\end{center}
\end{table}
 
\end{document}